\documentclass[twocolumn]{aastex702}

\usepackage{natbib}
\usepackage{lineno}
\usepackage{multirow}
\usepackage{booktabs}
\usepackage{epsfig}
\usepackage{ulem}
\usepackage{rotating}
\usepackage{graphicx}
\usepackage{url}
\usepackage{xcolor}
\usepackage{amsmath}
\usepackage{color}
\usepackage{savesym}
\savesymbol{tablenum}
\usepackage{siunitx}
\usepackage[normalem]{ulem}
\usepackage{appendix}
\usepackage{enumerate}
\usepackage{gensymb}
\usepackage{longtable}
\usepackage{afterpage}
\usepackage{threeparttable}
\usepackage{tabularx}
\usepackage[bottom]{footmisc}

\DeclareSIUnit\parsec{pc}
\DeclareSIUnit\h{\textit{h}}

\begin{document}
    \title{Constraining Intergalactic Dust with Supernova Color Evolution: Current Constraints from DES and Forecasts for LSST and \textit{Roman}}
    \author[0000-0001-8596-4746]{Erik R.~Peterson}\email{erikpete@umich.edu}
    \affiliation{Department of Physics, University of Michigan, Ann Arbor, MI 48109, USA}
    \affiliation{Society of Fellows, University of Michigan, Ann Arbor, MI 48109, USA}
    \author[0000-0001-5201-8374]{Dillon Brout}\email{}
    \affiliation{Department of Astronomy, Boston University, Boston, MA 02215, USA}
    \affiliation{Department of Physics, Boston University, Boston, MA 02215, USA}
    \author[0000-0002-4934-5849]{Daniel Scolnic}\email{}
    \affiliation{Department of Physics, Duke University, Durham, NC 27708, USA}
    \author[0000-0002-7234-844X]{Bastien Carreres}\email{}
    \affiliation{Department of Physics, Duke University, Durham, NC 27708, USA}
    \author[0000-0003-3917-0966]{Rebecca C.~Chen}\email{}
    \affiliation{Kavli Institute for Particle Astrophysics \& Cosmology, Stanford University, Stanford, CA 94305, USA}
    \affiliation{SLAC National Accelerator Laboratory, Menlo Park, CA 94025, USA}
    \affiliation{Department of Physics, Stanford University, Stanford, CA 94305, USA}
    \author[0000-0003-3221-0419]{Richard Kessler}\email{}
    \affiliation{Kavli Institute for Cosmological Physics, University of Chicago, Chicago, IL 60637, USA}
    \affiliation{Department of Astronomy and Astrophysics, University of Chicago, Chicago, IL 60637, USA}
    \author[0000-0002-8012-6978]{Brodie Popovic}\email{}
    \affiliation{School of Physics and Astronomy, University of Southampton, Southampton, SO17 1BJ, UK}
    \author[0000-0002-1873-8973]{Benjamin M. Rose}\email{}
    \affiliation{Department of Physics, Baylor University, Waco, TX 76798, USA}
    \author[0000-0001-8788-1688]{Maria Vincenzi}\email{}
    \affiliation{Department of Physics, University of Oxford, 
    Oxford, OX1 3RH, UK}
    \author[0000-0002-5389-7961]{Maria Acevedo}\email{}
    \affiliation{Department of Physics, Duke University, Durham, NC 27708, USA}
    \author[0000-0002-2495-3514]{John Franklin Crenshaw}\email{}
    \affiliation{Kavli Institute for Particle Astrophysics \& Cosmology, Stanford University, Stanford, CA 94305, USA}
    \affiliation{SLAC National Accelerator Laboratory, Menlo Park, CA 94025, USA}
    \affiliation{Department of Physics, Stanford University, Stanford, CA 94305, USA}
    \author[0000-0002-1296-6887]{Lluís Galbany}\email{}
    \affiliation{Institute of Space Sciences (ICE-CSIC), Campus UAB, Barcelona, 08193, Spain}
    \affiliation{Institut d’Estudis Espacials de Catalunya (IEEC), Castelldefels, Barcelona, 08860, Spain}
    \author[0000-0002-6741-983X]{Matthew Grayling}\email{}
    \affiliation{Institute of Astronomy and Kavli Institute for Cosmology, 
    University of Cambridge,
    Cambridge, CB3 0HA, UK}
    \author[0000-0002-4884-6756]{Benne W.~Holwerda}\email{}
    \affiliation{Department of Physics and Astronomy, University of Louisville, 
    Louisville, KY 40292, USA}
    \author[0000-0002-9883-7460]{Jacqueline McCleary}\email{}
    \affiliation{Department of Physics, Northeastern University, Boston, MA 02115, USA}
    \author[0000-0002-9436-8871]{Ayan Mitra}\email{}
    \affiliation{Center for AstroPhysical Surveys, National Center for Supercomputing Applications, Urbana, IL 61801, USA}
    \affiliation{Department of Astronomy, University of Illinois at Urbana-Champaign, Urbana, IL 61801, USA}
    \author[0009-0003-6728-9291]{Rujuta Purohit}\email{}
    \affiliation{Department of Astronomy, Boston University, Boston, MA 02215, USA}
    \author[0000-0002-3500-6635]{Corentin Ravoux}\email{}
    \affiliation{Université Clermont-Auvergne, CNRS, LPCA, Clermont-Ferrand, 63000, France}
    \author[0000-0003-2764-7093]{Masao Sako}\email{}
    \affiliation{Department of Physics and Astronomy, University of Pennsylvania, Philadelphia, PA 19104, USA}
    \author[0000-0002-8687-0669]{Bruno O.~Sánchez}\email{}
    \affiliation{Centre de Physique des Particules de Marseille, CNRS, Marseille, 13288, France}
    \author[0000-0003-0347-1724]{Christopher W.~Stubbs}\email{}
    \affiliation{Harvard University, Department of Physics,
    Cambridge, MA 02138, USA}
    \affiliation{Center for Astrophysics, Harvard \& Smithsonian, 
    Cambridge, MA 02138, USA}
    \author[0000-0002-9413-4186]{Syed Uddin}\email{}
    \affiliation{American Public University System, 
    Charles Town, WV 25414, USA}
    \affiliation{Center for Astronomy, Space Science and Astrophysics, Independent University, Bangladesh, Dhaka 1245, Bangladesh}
    \author[]{The LSST Dark Energy Science Collaboration}\email{}
    \affiliation{SLAC National Accelerator Laboratory, Menlo Park, CA 94025, USA}
    \author[]{The Roman Science Collaboration}\email{}
    \affiliation{NASA Goddard Space Flight Center,
    Greenbelt, MD 20771, USA}

\begin{abstract}

Dust is an important systematic in precision Type Ia supernova (SN Ia) cosmology, and while dust in the Milky Way and SN host galaxies has been extensively studied, the impact of potential \textit{intergalactic dust} is less well constrained. Using estimates from \citet{Menard10}, we find that current intergalactic dust bounds could shift SN distance moduli by up to 0.02 mag at $z \gtrsim 1$ from simulations.
However, we show dust models can be constrained independently by SN samples themselves via a specific predicted redshift-dependent evolution observed in SN colors. 
While current data from the Dark Energy Survey (DES) cannot distinguish between various amounts of intergalactic dust in our universe, we demonstrate that the Roman Space Telescope (\textit{Roman}) will place a significant constraint on intergalactic dust models ($>$ 3$\sigma$).
In contrast, the Rubin Observatory's Legacy Survey of Space and Time (LSST), even with its unprecedented statistics, will only reach $\lesssim$~2$\sigma$ significance on any given intergalactic dust prediction relative to no dust.
We propagate multiple parameterizations of intergalactic dust through to cosmology.
For LSST, we see a $\sim$60\% reduction in the figure of merit (FoM) for $w$CDM and a $\sim$20\% reduction for $w_0w_a$CDM, and for \textit{Roman}, a $\sim$20\% reduction in the FoM for $w$CDM and a $\sim$5\% reduction for $w_0w_a$CDM. 
Utilizing \textit{Roman} as an external constraint on dust for LSST results in a slightly improved FoM ($\sim$50\% reduction) for $w$CDM for LSST.
Finally, we show that intergalactic dust uncertainties decrease the FoM of the previous DES analysis (16.9\% for $w$CDM and 8.1\% for $w_0w_a$CDM) and that intergalactic dust cannot explain away recent hints for evolving dark energy.

\end{abstract}

\section{Introduction}

The recent Dark Energy Spectroscopic Instrument (DESI) analyses \citep[e.g.,][]{DESI24,DESI25} have made it clear that, depending on the analysis and treatment of systematics, similar Type Ia supernova (SN Ia) samples such as Pantheon+ \citep{Brout22,Scolnic22} and Union3 \citep{RubinUNITY} can result in different uncertainties for cosmological parameter measurements.
One major difference between these two analyses is that Union3 incorporated \textit{intergalactic dust} as a systematic and found it to be the second most relevant systematic when measuring the energy density of matter, $\Omega_\textrm{M}$ \citep[table 6 of][]{RubinUNITY}. Intergalactic dust impacts how dust as well as SN color evolves as a function of redshift.
To further investigate the extent to which intergalactic dust impacts cosmological parameters, we analyze how well the Dark Energy Survey \citep[DES;][]{DES24} can constrain this evolution as a function of redshift. Additionally, we forecast how this systematic will impact the Vera C.~Rubin Observatory's Legacy Survey of Space and Time \citep[LSST;][]{LSST_DC2,Sanchez22} as well as the Nancy Grace Roman Space Telescope \citep[\textit{Roman};][]{Spergel15}.

Most SN Ia cosmology analyses attempt to correct for the impacts from \textit{interstellar} dust in both the Milky Way and in the galaxies hosting SNe without considering \textit{intergalactic} dust \citep[e.g.,][]{Brout22,DES24,Popovic24}.
Recognizing that the observed SN color distribution is determined by the intrinsic SN color distribution and affected by the intervening dust, SN color is often parameterized by: 
\begin{equation}
    c_\textrm{obs} = c_\textrm{int} + E_\textrm{dust} + \epsilon_\textrm{noise},
\end{equation}
\noindent where $c_\textrm{obs}$ and $c_\textrm{int}$ are the observed and intrinsic SN colors, respectively; $E_\textrm{dust}$ is the amount of dust extinction; and $\epsilon_\textrm{noise}$ is unaccounted for measurement noise \citep{Mandel11,Mandel17,BroutScolnic21,Popovic23}.
Teasing out the respective contributions to the observed SN color from the intrinsic color distribution versus effects from dust in host galaxies is difficult; however, given that dust can only impact color by reddening the light and doing so in a well-defined manner, with enough observations across a wide enough wavelength range, we can disentangle the two effects and measure the amount of intervening dust as well as the underlying intrinsic SN color distribution.

The impact from intergalactic dust is understood to increase as a function of redshift (linked to the number of intervening line-of-sight dark-matter halos) where we observe that the integrated column density rises with $z$ \citep[e.g.,][]{Menard10}. But, it does not increase linearly with distance, as the comoving dust density was lower at early times \citep[e.g.,][]{Menard08,Menard10,Amanullah10}.
\citet{RubinUNITY}, who find intergalactic dust to be the second largest systematic for $\Omega_\textrm{M}$, estimate the impact from intergalactic dust for their analysis using the number density of galaxies estimated from \citet{Menard10} (hereafter M10), and scaling the impact from dust following \citet{Menard08} and \citet{Amanullah10}. Unlike interstellar dust in the Milky Way or in each respective host galaxy, intergalactic dust should produce a characteristic evolution with redshift that can be modeled and measured. Once we are able to accurately measure the distribution of intergalactic dust in the universe, we can quantify its impact on cosmological parameters.

In Section~\ref{sec:data}, we describe the real data we work with from DES as well as simulated data for both LSST and \textit{Roman}.
In Section~\ref{sec:simulations}, we detail the simulations that we construct for all three surveys both including intergalactic dust and without. 
In Sections~\ref{sec:methods} and~\ref{sec:results}, we lay out our methods for distinguishing between differing amounts of intergalactic dust with SNe Ia and share our results on their relative impacts to cosmology.
And finally, in Sections~\ref{sec:discussions} and~\ref{sec:conclusions}, we provide our discussions and conclusions for this work.

\section{Data}\label{sec:data}
In order to test our ability to constrain intergalactic dust and understand it as a systematic, we use real SN data from DES as well as simulated data from both LSST and \textit{Roman} for our analysis and describe each survey below.

\subsection{DES}

Light curves (LCs) from DES come from \citet{Sanchez24}.
The DES sample consists of about 1,630 SN LCs passing cosmology-grade quality cuts which are photometrically classified and were observed between 2013 and 2018 with the Dark Energy Camera \citep[DECam;][]{Flaugher15} at the Victor M.~Blanco telescope on Cerro Tololo.
The general cadence for the DES LCs is roughly 1 week, and observations were taken in optical \textit{griz} bands.
The final DES cosmology sample spans a redshift range of $0.10<z<1.13$.
Since the DES5YR analysis \citep{DES24,Vincenzi24}, which we follow here, \citet{Popovic25} provided an updated calibration for DES, but this does not impact our SN color evolution results because we only use the SN color from the LCs.

The DES survey covered ten fields, each covering approximately 3 deg$^2$. The two deep fields from DES, designated ``C3'' and ``X3,'' reached a single-epoch depth of $\sim$24.5 mag in \textit{r}-band and were centered around right ascension/declination 3:30:35.6/$-$28:06:00.0 and 2:25:48.0/$-$4:36:00.0, respectively \citep{Smith20DES}.

\subsection{LSST}

LSST \citep{LSST_DC2,Sanchez22} has just begun and will span ten years of data collection with unprecedented scale. 
The Rubin Observatory is located on Cerro Pachón in Chile.
The telescope itself, the Simonyi Survey Telescope, has an 8.4 meter primary mirror, a 9.6 deg$^2$ field of view, and \textit{ugrizy} bandpass filters.\footnote{\url{https://rubinobservatory.org/for-scientists/rubin-101/key-numbers}.}
The survey will have a standard 30-second exposure time per visit, with single visit depths of $\sim$23.5--24.5 mag in optical bands.
LSST is expected to observe hundreds of thousands of SNe out to $z\sim1.2$ through its Wide Fast Deep (WFD) and Deep Drilling Field (DDF) programs.

The WFD survey covers approximately 19,600 deg$^2$, and, after year one, will employ a rolling cadence where specific fields will be observed at higher and lower cadences every other year.\footnote{\url{https://survey-strategy.lsst.io/baseline/wfd.html}.}
LSST has five dedicated deep fields for its DDF program that overlap with historically covered deep fields and will each receive on the order of 20,000 visits.\footnote{\url{https://survey-strategy.lsst.io/baseline/ddf.html}.}

\subsection{Roman}

\textit{Roman} \citep{Spergel15} is a NASA mission which launched at the end of August 2026 and will observe SNe Ia out to $z\sim3$.
\textit{Roman} will focus observations in the near-infrared (and at higher redshifts, thereby probing rest-frame optical light) with its 2.4 meter primary mirror and Wide Field Instrument (WFI; field of view of $\sim$0.28 deg$^2$).
According to the current survey strategy, the High-Latitude Time Domain Survey \citep[HLTDS;][]{Rose21Roman,Hounsell23,Rose25} will include both a shallower $\sim$19-deg$^2$ field in \textit{RZYJ} bands (formally F062, F087, F106, and F129) to a depth of $\sim$25.5 mag as well as a deeper $\sim$4-deg$^2$ field in \textit{YJHF} bands (formally F106, F129, F158, and F184) to a depth of $\sim$26.5 mag each with a 5-day cadence.
\citet{Aldoroty26} demonstrate that the survey goal of flux precision $<$1\% for photometry is achievable and that the deep fields are expected to have a detection efficiency of 50\% out to $\sim$26 mag in almost every band.

\begin{figure*}[!t]
    \centering
    \includegraphics[width=\textwidth]{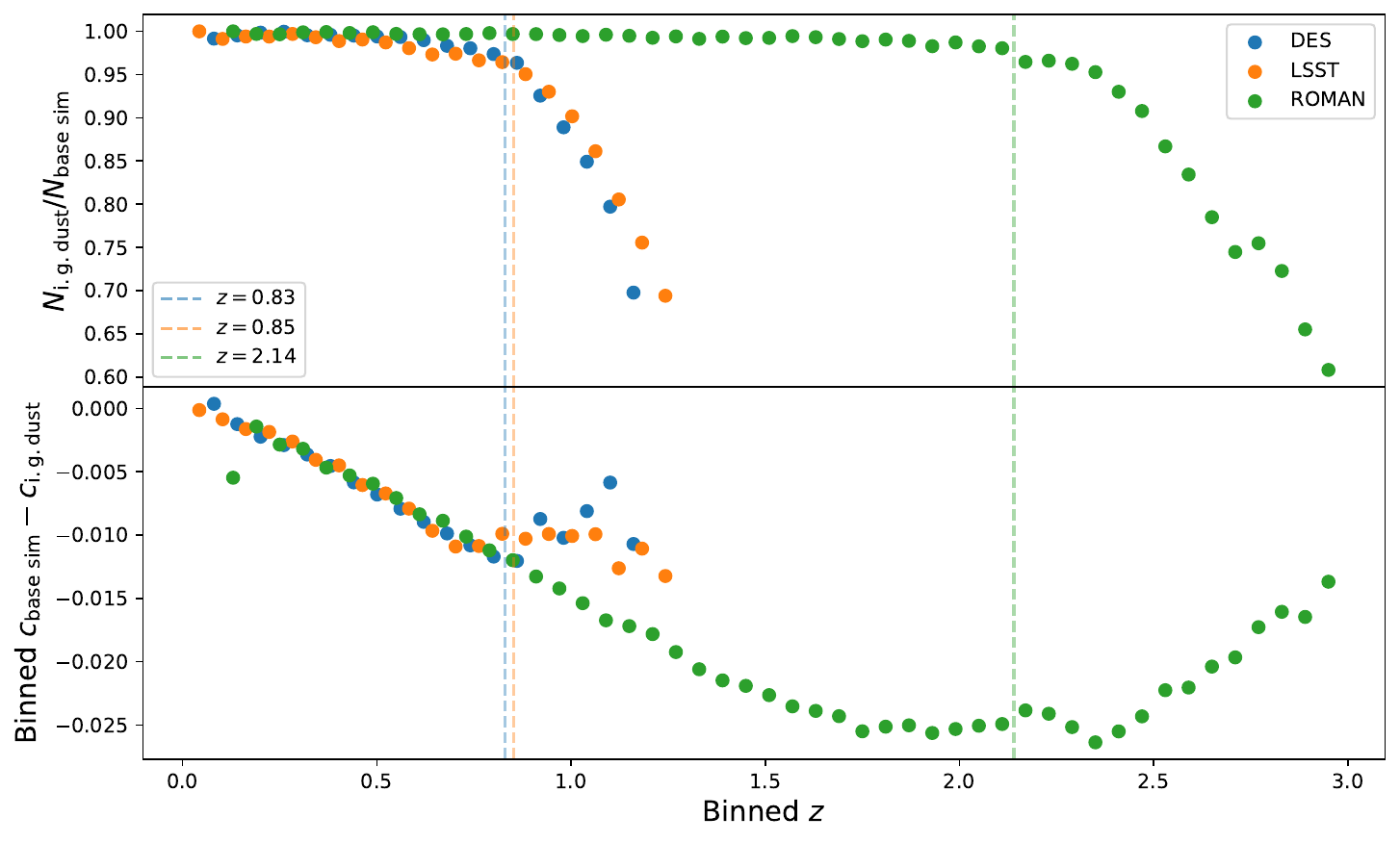}
    \caption{\textbf{Upper Panel}: Simulated ratio of binned SN counts as a function of redshift between our nominal intergalactic dust simulation (M10$_\textrm{max}$) and our baseline simulation with no intergalactic dust for the deep fields. Vertical lines are provided when selection effects grow large enough where more than a 1\% drop in the ratio of SN counts is observed for each simulation.
    \textbf{Lower Panel}: Binned differences in median colors between intergalactic dust and no intergalactic dust as a function of redshift for the deep fields.}
    \label{fig:DES_sim_compare}
\end{figure*}

\section{Simulations}\label{sec:simulations}

We construct simulations using the SN analysis package, \texttt{SNANA} \citep{Kessler09,Kessler19}.
All of our simulations have the same underlying intrinsic scatter model from \citet{Popovic23} which follows the framework from \citet{BroutScolnic21}.
The simulations derive characteristics from the data such as the LC coverage and cadence.
Specifically, the DES simulations are constructed following \citet{Vincenzi24}.
The simulated dust distribution characteristics for parameters such as $R_V$ and $\tau_E$ 
are taken from table~3 of \citet{Vincenzi24}.\footnote{The column labeled P21($M_*$) in \citet{Vincenzi24}.}

For both the LSST and \textit{Roman} simulations, we simulate each survey as laid out by their respective observing strategies. 
For the LSST and \textit{Roman} survey-sized simulations, which we call ``Sim Data,'' we specifically do not include intergalactic dust, as this is our baseline simulation.
Summary statistics from each of these simulations after applying typical quality cuts\footnote{Quality cuts for all samples include: $|x_1|<3$, $|c|<0.3$, $\sigma_{x_1}<1$, $\sigma_{t_\textrm{peak}}<2$ days, and LC fit probability from \texttt{SNANA} $>0.001$.} are provided in Table~\ref{tab:sim_characteristics}.
We provide counts for both our real (in the case of DES) and simulated (for both LSST and \textit{Roman}) data samples (N$_\textrm{data}$) as well as the much larger simulated samples used for bias corrections (N$_\textrm{biascor}$).
Following \citet{Vincenzi24}, we implement 7-dimensional bias corrections for all of our simulations which account for biases due to redshift, light-curve color (at the individual-SN and survey level), light-curve stretch (also at the individual-SN and survey level), host parameters, and intrinsic scatter. 
BEAMS with Bias Corrections \citep[BBC;][]{KesslerScolnic17} are carried out by simulating a much larger ``biascor'' sample that fills the complete parameter space for each of the desired parameters and estimating the cosmological biases introduced by not filling out the complete parameter space with the data.

\begin{table}[t]
\caption{Characteristics of both the data-sized and much larger simulations we construct for all three surveys after cuts overall and in their respective deep fields. For DES, we use real data.}\label{tab:sim_characteristics}
\begin{threeparttable}
\begin{tabularx}{\columnwidth}{l@{\extracolsep{\fill}}cccl}
\hline \hline
Survey & Field & N$_{\textrm{data}}$ &  N$_{\textrm{biascor}}$ & Redshift Range \\
\hline
\multirow{2}*{DES} & Total & 1,627 & 354,102 & \multirow{2}*{$0.05 < z < 1.20$\textsuperscript{a}} \\
& Deep & 494 & 130,902 &\\
\hline
\multirow{2}*{LSST} & Total & 138,500 & 5,541,224 & \multirow{2}*{$0.01 < z < 1.29$}\\
& Deep & 3,646 & 143,486 & \\
\hline
\multirow{2}*{\textit{Roman}} & Total & 12,654 & 508,177 & \multirow{2}*{$0.10 < z < 3.00$}\\
& Deep & 3,799 & 152,811 &\\
\hline
\end{tabularx} 
\begin{tablenotes}
\item[a] {The biascor simulations span a larger redshift range than the final DES cosmology sample.}
\end{tablenotes}
\end{threeparttable}
\end{table}

All simulations for a given survey have the same underlying characteristics. Simulations with added intergalactic dust are carried out in \texttt{SNANA} by adding a redshift-dependent extinction, $A_V(z)$, following the intergalactic dust curves derived in M10. In \texttt{SNANA} this is implemented using the \texttt{ZVARIATION\_POLY} function and modulating the \texttt{EBV\_IGM} parameter with the desired polynomial coefficients to describe a given intergalactic dust parameterization. Coefficients used in this analysis for all three intergalactic dust curves in figure~9 of M10, which we call M10$_\textrm{max}$, M10$_\textrm{med}$, and M10$_\textrm{min}$ (dotted cyan, solid cyan, and solid dark blue, respectively), are provided in Appendix~\ref{appendix:M10_reproduction}.

In order to better characterize the differences between our simulations both with (following M10$_\textrm{max}$) and without intergalactic dust (our baseline simulation), we provide comparative binned statistics between the two in Fig.~\ref{fig:DES_sim_compare}.
Specifically, we plot the ratio of SN counts with and without intergalactic dust (upper panel) and difference in median color in redshift bins (lower panel).
In the upper panel, we also indicate redshifts where the selection effects grow large enough for there to be a 1\% drop for the intergalactic dust simulations as compared to the simulations without intergalactic dust for each survey.
We use these redshifts as our upper limits for our analysis when trying to constrain intergalactic dust from SN color evolution in Section~\ref{sec:results}.
The significant drop in the ratio of SN counts in both the shallow and deep fields at higher redshifts is expected as added dust (which increases as a function of redshift) will result in fainter SNe that will be selected out more frequently.
In the lower panel of Fig.~\ref{fig:DES_sim_compare}, we observe that, according to our simulations, in both the shallow and deep fields we expect to observe a gradual trend towards redder median $c$ values with the inclusion of intergalactic dust as a function of redshift until selection effects take over.
This too is expected, as the only difference between the two simulations is added dust, therefore, the simulation with added dust will be both more extincted and redder as a function of redshift.
In this lower panel, we also observe the impact from selection effects whereby the consistent trends in color break when selection effects begin to significantly impact the intergalactic dust simulation.
This further justifies using the redshifts indicated in Fig.~\ref{fig:DES_sim_compare} as our upper limits when analyzing SN color evolution to constrain intergalactic dust later on in this work.

\section{Method}\label{sec:methods}

In order to observe the effect from intergalactic dust, we observe the SN color evolution as a function of redshift.
Intergalactic dust should have a characteristic effect that results in maintaining a redder mean color as a function of redshift until selection effects take over.
Therefore, on a SN Ia color versus redshift diagram, one would expect such a curve with intergalactic dust to remain flatter, or even trend redder, for longer than a realization without intergalactic dust.

In the SALT model framework \citep{Guy07,Kenworthy21}, a light-curve stretch, $x_1$, light-curve normalization factor, $x_0$, and light-curve color, $c$, are fit to the SN Ia photometry in order to standardize them.
Combining with the BBC method \citep{KesslerScolnic17}, distances are obtained following a modified version of the Tripp equation \citep{Tripp98}:
\begin{equation}
    \mu = m_B + \alpha x_1 - \beta c + \gamma + \delta_{\mu-\textrm{bias}} - \mathcal{M},
\end{equation}
\noindent where $\alpha$ and $\beta$ are globally fit nuisance parameters relating $x_1$ and $c$, respectively, to luminosity corrections; the peak magnitude, $m_B$, is $-2.5\log(x_0)$ plus a constant; $\gamma$ is the host galaxy mass-luminosity correction \citep[e.g.,][]{Sullivan10}; $\delta_{\mu-\textrm{bias}}$ is the bias correction derived from the BBC method; and $\mathcal{M}$ is the absolute magnitude of a SN Ia with $x_1 = c = 0$.
We use these fitted distances for our cosmological analysis later on in Section~\ref{subsec:cosmo_analysis}.

To compare simulations with and without intergalactic dust, we take the fitted SN colors and bin them as a function of redshift in order to study their SN color evolutions and observe their differences.
We note that in the SALT model framework the model's reference color is defined so that the mean color of the training sample is approximately $c=0$. 
Because of this, when comparing the SN color evolutions from the simulations to the data (or simulated data in the cases of LSST and \textit{Roman}), this test is insensitive to a color offset between data and simulation and is only sensitive to the shape of the curve.
Consequently, we fit for an offset between the data and simulated mean color values using a weighted average and subtract out the offset.
Comparing these resulting SN color evolutions from the simulations both with and without intergalactic dust to the (simulated) data, we report their $\chi^2$ values defined as:
\begin{equation}
    \chi^2 = \sum \frac{ (\bar c_\mathrm{data} - \bar c_\mathrm{sim})^2}{\sigma_{\bar c}^2},
\end{equation}
\noindent where $\bar c$ is the mean color in each redshift bin from the data and simulation, respectively, and $\sigma_{\bar c}$ is the standard error from the data in that bin.

\begin{figure}
    \centering
    \includegraphics[width=\columnwidth]{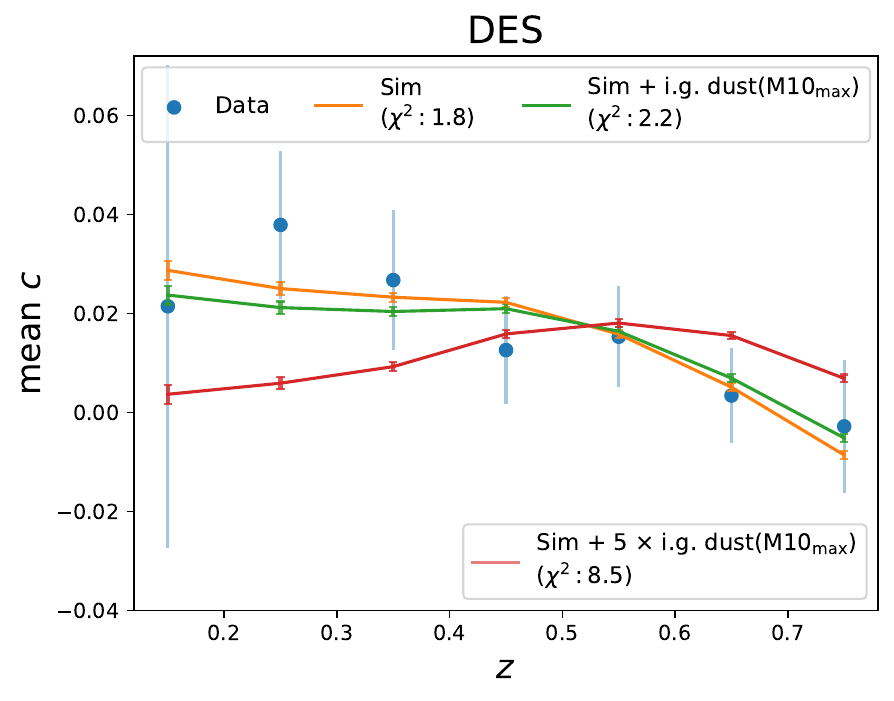}
    \caption{Binned mean color values as a function of redshift for both real DES data and simulations in the DES deep fields. 
    Three sets of simulations are provided, one without intergalactic dust included (orange), one with the nominal amount of intergalactic dust used in this analysis included (green), and one with five times the nominal amount of intergalactic dust included (red). DES data are indicated in blue, and standard error values in each bin are provided for both the data and the simulations. 
    }
    \label{fig:c_evol_DES}
\end{figure}

\section{Results}\label{sec:results}

First, we analyze to what extent we can constrain the amount of intergalactic dust there is using real data from DES. Then, we forecast to what extent we will be able to constrain the amount of intergalactic dust with both LSST and \textit{Roman}.
Subsequently, we estimate the impact intergalactic dust has as a systematic on cosmology for all of these surveys.

\subsection{Constraining Intergalactic Dust with Real Data}

By plotting binned mean color values for both simulations and data as a function of redshift in Fig.~\ref{fig:c_evol_DES}, we can observe which simulation is preferred by the data.
To explore how well current data can constrain intergalactic dust, as a first test, we use DES to place an upper limit on intergalactic dust.
As such, we take the extinction curve from M10 with the largest impact from intergalactic dust provided (their figure~9, dotted cyan curve, which we call M10$_\textrm{max}$; see Appendix~\ref{appendix:M10_reproduction}) for our nominal intergalactic dust simulation for our analysis with DES.
With this implementation of intergalactic dust, in 
Fig.~\ref{fig:c_evol_DES}, we observe that neither the simulation with intergalactic dust nor without is clearly preferred by the data; the direct $\chi^2$ value with intergalactic dust is 2.2 and without is 1.8.

In Fig.~\ref{fig:c_evol_DES}, we include an additional implementation of intergalactic dust, but with five times the nominal amount of intergalactic dust added ($5\times \textrm{M10}_\textrm{max}$).
As can be seen from Fig.~\ref{fig:c_evol_DES}, the trend in color observed from the intergalactic dust simulation with extra intergalactic dust goes as expected, where with added dust, the mean color remains redder as a function of redshift for longer as opposed to trending bluer more quickly due to selection effects.
In terms of constraining intergalactic dust, $5\times \textrm{M10}_\textrm{max}$ can likely be ruled out given its $\chi^2$ value of 8.5 and $\Delta\chi^2$ of 6.7 as compared to no intergalactic dust; this translates to a significance of $\sim$2.6$\sigma$.

\begin{figure}
    \centering
    \includegraphics[width=\columnwidth]{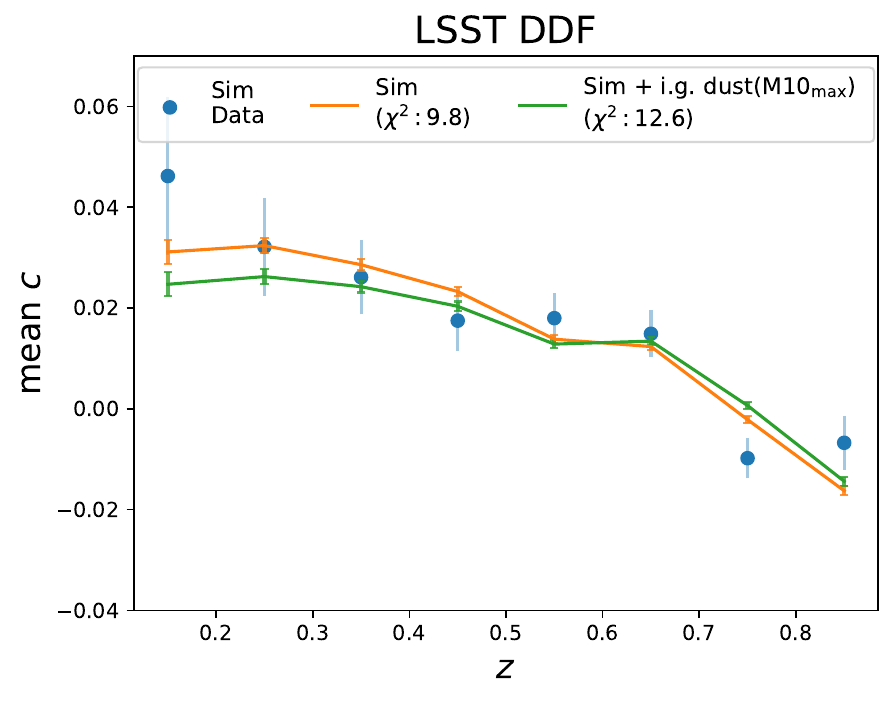}

    \caption{Similar to Fig.~\ref{fig:c_evol_DES}, but binned mean color values as a function of redshift comparing LSST simulated data (i.e., expected data) to LSST simulations in the LSST deep fields.}
    \label{fig:c_evol_LSST}

\end{figure}

\begin{figure}[!t]
    \centering
    \includegraphics[width=\columnwidth]{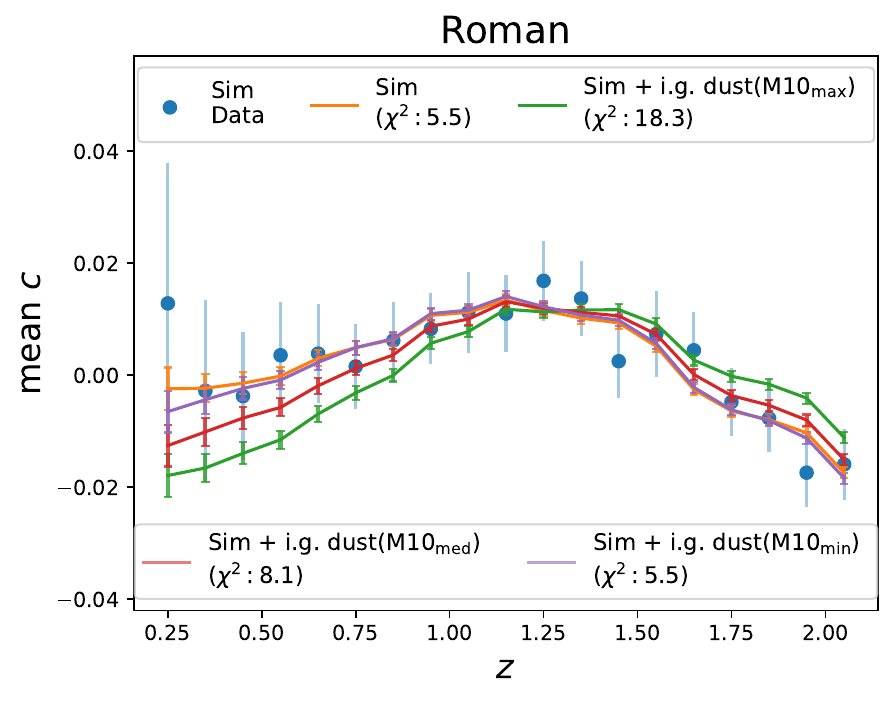}
    \caption{Similar to Figs.~\ref{fig:c_evol_DES} and~\ref{fig:c_evol_LSST}, but binned mean color values as a function of redshift for \textit{Roman} simulated data (i.e., expected data) as well as \textit{Roman} simulations in the \textit{Roman} deep fields. Additionally, rather than solely including the upper limit intergalactic dust curve from M10, here we additionally plot the lower limit and intermediate curves, which we label M10$_\textrm{min}$ and M10$_\textrm{med}$, respectively.}
    \label{fig:c_evol_ROMAN}

\end{figure}

\begin{figure*}[!t]
    \centering
    \includegraphics[width=\textwidth]{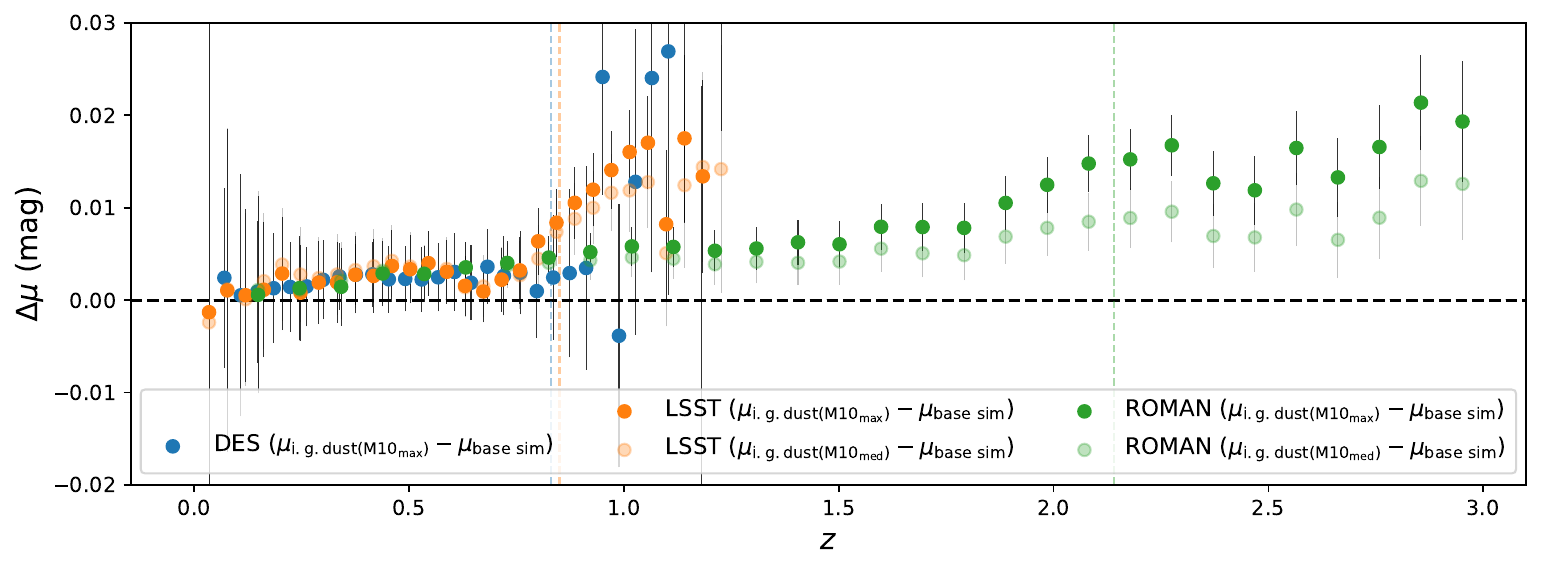}
    \caption{Binned differences in distance modulus values when including intergalactic dust and when not as a function of redshift. Standard error values are provided as error bars. Additional transparent points are provided for both LSST and \textit{Roman} given our potential ability to better distinguish between intergalactic dust models (in this case, M10$_\textrm{med}$) using SN color evolution by the time these surveys are complete. Vertical lines are also included indicating the redshift limit used for our analysis of SN color evolution (but not in our cosmological analysis) for each survey.}
    \label{fig:mu_from_igdust}
\end{figure*}

\subsection{Forecasted Constraints from Future Surveys}
\subsubsection{LSST Constraints}

Although LSST does not probe significantly deeper than DES did, with its unprecedented statistics, there are expectations that LSST will constrain intergalactic dust much better than DES.
This can be observed in the color evolution of SNe expected to be observed over the course of the full LSST survey in the DDF fields in Fig.~\ref{fig:c_evol_LSST}, where the uncertainties on the mean color in each redshift bin (on the order of $\sim$0.01) are much smaller than those from DES (on the order of $\sim$0.02).

The difference in $\chi^2$ values between the simulations with intergalactic dust (specifically M10$_\textrm{max}$) and without intergalactic dust for DES is 0.4. 
As expected, LSST will be better equipped to differentiate between the two sets of simulations more significantly, with $\chi^2$ values of 12.6 and 9.8 for the intergalactic dust simulation and no intergalactic dust simulation, respectively, as compared to simulated (expected) data from LSST with no intergalactic dust included. 
That being said, this $\Delta \chi^2$ of 2.8 is only $\sim$1.7$\sigma$.
As a result, we determine that LSST alone will be unable to provide a significant enough constraint on intergalactic dust and will need such a constraint from \textit{Roman} to limit this systematic.

Additional SN evolution curves, including other curves from M10 with less intergalactic dust as well as constraints from the shallow, WFD fields are provided in Appendix~\ref{appendix:extra_c_vs_z}.
We find that although there truly are unprecedented statistics in the WFD fields, given the limited redshift range covered, the simulation with intergalactic dust is largely indistinguishable from the simulation with no intergalactic dust in this regime.

\subsubsection{Roman Constraints}
In a similar manner to our DES and LSST analyses, we aim to forecast how well \textit{Roman}, which will reach much higher in redshift, will be able to constrain intergalactic dust.
Rather than setting an upper limit on the amount of intergalactic dust that can be ruled out with real data though, here, we predict a lower limit on the amount of dust for which \textit{Roman} will be able to detect a signal.
To do so, we not only include the maximum curve from M10, M10$_\textrm{max}$, but also take both the intermediate predicted curve (their figure~9, solid cyan curve, which we call M10$_\textrm{med}$; see Appendix~\ref{appendix:M10_reproduction}) and smallest predicted curve from M10 (their figure~9, solid dark blue curve, which we call M10$_\textrm{min}$; see Appendix~\ref{appendix:M10_reproduction}) and try to see if \textit{Roman} can distinguish between these amounts of intergalactic dust and no intergalactic dust in Fig.~\ref{fig:c_evol_ROMAN}.

In Fig.~\ref{fig:c_evol_ROMAN} we compare the SN color evolution from these four simulations, one without intergalactic dust, our baseline simulation, as well as three with intergalactic dust (the M10$_\textrm{min}$, M10$_\textrm{med}$, and M10$_\textrm{max}$ implementations of intergalactic dust).
The ``Sim Data" we simulate is a simulated sample that is expected from \textit{Roman} (in this instance without intergalactic dust).
When comparing this simulated data to our simulations with and without intergalactic dust in Fig.~\ref{fig:c_evol_ROMAN}, we observe that \textit{Roman} will have the ability to distinguish between no intergalactic dust ($\chi^2=5.5$) and the maximum intergalactic dust curve provided by M10 ($\chi^2=18.3$) at a significance of $\sim$3.6$\sigma$.
Distinguishing between even smaller amounts of intergalactic dust, however, may be more challenging, as M10$_\textrm{med}$ and M10$_\textrm{min}$ result in $\Delta \chi^2$ values of 2.6 and 0.0, respectively.

\begin{table*}[t]
\caption{Cosmological constraints with Flat-$\Lambda$CDM.}\label{tab:flcdm}
\begin{threeparttable}
\begin{tabularx}{\textwidth}{l@{\extracolsep{\fill}}ccccr}
\hline \hline\\[-3mm]
Survey & Dust Constraint & Cosmo.~Constraint & $\Delta \Omega_\textrm{M}$ & $\sigma_{\delta \Omega_\textrm{M}}$ & $\frac{\sigma_{\delta \Omega_\textrm{M}}}{\sigma_{\textrm{stat},\Omega_\textrm{M}}}$\\[2mm]
\hline
\multirow{2}*{DES} & i.g.~dust(M10$_\textrm{max}$) & \multirow{2}*{SN only} & 0.0114 & 0.0104 & 86.3\%\\
& i.g.~dust(M10$_\textrm{max}$) $\times$ 5 & & 0.0161 & 0.0216 & 180.5\%\\
\hline
\multirow{2}*{LSST 1yr} & i.g.~dust(M10$_\textrm{med}$) & \multirow{2}*{SN only} & 0.0047 & 0.0063 & 177.4\%\\
 & i.g.~dust(M10$_\textrm{max}$) & & 0.0144 & 0.0065 & 182.6\%\\
\hline
\multirow{2}*{LSST 10yr} & i.g.~dust(M10$_\textrm{med}$) & \multirow{2}*{SN only} & 0.0002 & 0.0010 & 28.4\%\\
 & i.g.~dust(M10$_\textrm{max}$) & & 0.0092 & 0.0008 & 23.9\%\\
\hline
\multirow{2}*{\textit{Roman}} & i.g.~dust(M10$_\textrm{med}$) & \multirow{2}*{SN only} & 0.0000 & 0.0020 & 50.7\%\\
 & i.g.~dust(M10$_\textrm{max}$) & & 0.0013 & 0.0038 & 94.9\%\\
\hline
\end{tabularx} 
\end{threeparttable}
\end{table*}

\subsection{Impact as a Systematic on Cosmology}\label{subsec:cosmo_analysis}

From simulations we observe that the impact from intergalactic dust on distances is not expected to be very large.
In Fig.~\ref{fig:mu_from_igdust}, we compare $\mu$ values when incorporating intergalactic dust versus when not in bins of redshift and find that no bin is more than 2$\sigma$ from zero impact for DES and no more than $\sim$3$\sigma$ for LSST at $z<1$ and \textit{Roman} at $z<2$.
We note a tail toward larger $\mu$ values when including intergalactic dust at the high redshift end (e.g., LSST reaches as much as 0.02 mag at $z$ $\approx$ 1), but we attribute this primarily to selection effects and find the error bars on those bins to be large.
From this figure, we may predict that intergalactic dust may not be highly impactful to cosmology.

To test this further, we run each of our simulations through to cosmology using \texttt{wfit} in \texttt{SNANA} with intergalactic dust as a systematic and report our results on Flat-$\Lambda$CDM, $w$CDM, and $w_0w_a$CDM in Tables~\ref{tab:flcdm},~\ref{tab:wcdm}, and~\ref{tab:w0wacdm}, respectively.
For both $w$CDM and $w_0w_a$CDM, cosmological constraints are given for both SNe only as well as with a CMB prior from \citet{Planck18}.
In each of the tables we report changes to the relevant cosmological parameters' central values and uncertainties when considering a given amount of intergalactic dust as opposed to no intergalactic dust.
We provide simple differences ($\Delta$; i.g.~dust $-$ no dust) which we also call biases; differences in uncertainty in quadrature ($\sigma_\delta$; which we report as negative when the uncertainty shrinks with i.g.~dust); as well as the differences in uncertainty in quadrature divided by the statistical uncertainty alone for that parameter ($\sigma_\delta/\sigma_\textrm{stat}$) for a measure of how big the change in systematic uncertainty is as compared to the statistical uncertainty.
Lastly, for both $w$CDM as well as $w_0w_a$CDM we provide an additional statistic, $\Delta$FoM/FoM, which is defined as the difference in figure of merit (FoM) between a given systematic and statistical uncertainty alone and normalized by the statistical-alone FoM.

As a note, when fitting cosmology with LSST, given its unprecedented statistics, we found that statistical uncertainties on $\mu$ reached so low in the medium redshift bins (0.4 $<$ \textit{z} $<$ 0.6; $\ll$10$^{-2}$ mag) that our cosmological results were biased significantly.
Because of this, and because we are not including a full systematic error budget for this analysis, we add an error floor per bin of 10$^{-2}$ mag (as this was the minimum error floor necessary to remove the cosmological bias) to our cosmological parameter fits for our LSST simulations. We provide further information in Appendix~\ref{appendix:low_muerr} and leave an analysis of this peculiarity to future work by the community.

Additionally, while our nominal LSST simulation spans the full ten years of the survey (LSST 10yr), in order to project results from LSST on a sooner time scale, we also include cosmological results from LSST after just one year (LSST 1yr) by scaling down our simulated data by a factor of ten.

\begin{table*}[t]
\caption{Cosmological constraints with \textit{w}CDM.}\label{tab:wcdm}
\begin{threeparttable}
\begin{tabularx}{\textwidth}{l@{\extracolsep{\fill}}ccccccccr}
\hline \hline\\[-3mm]
Survey & Dust Constraint & Cosmo.~Const. & $\Delta w$ & $\sigma_{\delta w}$ & $\frac{\sigma_{\delta w}}{\sigma_{\textrm{stat},w}}$ & $\Delta \Omega_\textrm{M}$ & $\sigma_{\delta \Omega_\textrm{M}}$ & $\frac{\sigma_{\delta \Omega_\textrm{M}}}{\sigma_{\textrm{stat},\Omega_\textrm{M}}}$ & $\frac{\Delta\textrm{FoM}}{\textrm{FoM}}$\\[2mm]
\hline
\multirow{2}*{DES} & i.g.~dust(M10$_\textrm{max}$) 
 & \multirow{2}*{SN only} & $-$0.0197 & 0.0449 & 32.7\% & 0.0204 & 0.0186 & 22.3\% & $-$16.9\%\\
 & i.g.~dust(M10$_\textrm{max}$) $\times$ 5
 & & $-$0.0143 & 0.0537 & 39.1\% & 0.0107 & 0.0517 & 61.9\% & $-$39.2\%\\
\hline
& \multirow{2}*{i.g.~dust(M10$_\textrm{med}$)} & CMB prior & 0.0088 & 0.0063 & 37.7\% & 0.0046 & 0.0032 & 61.7\% & $-$49.7\%\\
 LSST & & SN only & $-$0.0072 & 0.0107 & 26.3\% & 0.0080 & 0.0080 & 52.8\% & $-$51.1\%\\
 1yr & \multirow{2}*{i.g.~dust(M10$_\textrm{max}$)} & CMB prior & 0.0137 & 0.0064 & 38.4\% & 0.0107 & 0.0039 & 74.5\% & $-$52.4\%\\
 & & SN only & $-$0.0521 & 0.0268 & 65.6\% & 0.0333 & 0.0121 & 80.6\% & $-$59.8\%\\
\hline
& \multirow{2}*{i.g.~dust(M10$_\textrm{med}$)} & CMB prior & $-$0.0005 & $-$0.0009 & $-$6.6\% & $-$0.0013 & 0.0027 & 62.1\% & $-$51.3\%\\
 LSST & & SN only & 0.0045 & 0.0042 & 19.7\% & $-$0.0044 & 0.0051 & 72.7\% & $-$54.3\%\\
 10yr & \multirow{2}*{i.g.~dust(M10$_\textrm{max}$)} & CMB prior & $-$0.0027 & 0.0040 & 29.1\% & 0.0062 & 0.0036 & 81.8\% & $-$53.0\%\\
 & & SN only & $-$0.0213 & 0.0143 & 66.6\% & 0.0125 & 0.0075 & 106.1\% & $-$60.1\%\\
\hline
\multirow{4}*{\textit{Roman}} & \multirow{2}*{i.g.~dust(M10$_\textrm{med}$)} & CMB prior & 0.0001 & 0.0016 & 6.2\% & 0.0001 & 0.0021 & 53.9\% & $-$14.6\%\\
 & & SN only & 0.0010 & 0.0139 & 25.7\% & $-$0.0004 & 0.0043 & 61.2\% & $-$23.6\%\\
 & \multirow{2}*{i.g.~dust(M10$_\textrm{max}$)} & CMB prior & 0.0008 & 0.0029 & 11.1\% & 0.0012 & 0.0029 & 74.4\% & $-$19.3\%\\
 & & SN only & $-$0.0056 & 0.0234 & 43.4\% & 0.0014 & 0.0063 & 90.0\% & $-$32.6\%\\
\hline
\end{tabularx} 
\end{threeparttable}
\end{table*}

\subsubsection{Flat-$\Lambda$CDM}
For Flat-$\Lambda$CDM, in Table~\ref{tab:flcdm}, we observe that for DES, the nominal amount of intergalactic dust from M10 results in an 86.3\% increase in the systematic uncertainty for $\Omega_\textrm{M}$ relative to the statistical uncertainty. This increases as more dust is added up to 180.5\% for i.g.~dust(M10$_\textrm{max}$) $\times$ 5. 
For LSST, results after one year are significantly impacted, where the shift in $\Omega_\textrm{M}$ is comparable to that observed for DES, but since the statistical uncertainty is smaller, the relative increase in systematic uncertainty is $>$170\%.
The impact on LSST after ten years is largely mitigated, where we only observe a $\sim$25\% increase in the systematic uncertainty in $\Omega_\textrm{M}$ relative to statistical uncertainty.
Finally, for \textit{Roman}, the increase in systematic uncertainty in $\Omega_\textrm{M}$ is comparable to that of DES ($\sim$50--95\% depending on how much dust is added) and the bias introduced is negligible. 

For all three surveys, in most cases the biases in $\Omega_\textrm{M}$ introduced by added intergalactic dust are smaller than or at most comparable to the added uncertainty from this additional systematic. This is evidence that intergalactic dust does not add a significant bias to our recovered cosmology.
This is not necessarily the case for LSST with i.g.~dust(M10$_\textrm{max}$) where the shift in $\Omega_\textrm{M}$ is sizably larger than the increase added in quadrature in systematic uncertainty from introducing this amount of i.g.~dust, but we attribute this to the inflated statistical uncertainty due to the error floor discussed in Appendix~\ref{appendix:low_muerr}.

\subsubsection{$w$CDM}
When we extend to $w$CDM, in Table~\ref{tab:wcdm}, we see that for all surveys shifts in both $w$ and $\Omega_\textrm{M}$ are either subdominant to or comparable to the increased uncertainty due to each systematic variant (except for M10$_\textrm{max}$ with LSST as was discussed above).
DES results demonstrate large increases in systematic uncertainty relative to statistical uncertainty alone ($\sim$20--60\% increases for SN only).
These large increases in systematic uncertainty result in significant reductions in FoM with DES up to $\sim$40\% depending on how much intergalactic dust is truly present.

Intergalactic dust appears to be equally or perhaps even more detrimental to the FoM for LSST (both LSST 1yr and LSST 10yr) in $w$CDM as compared to DES. Regardless of which intergalactic dust variant is used and both with and without a CMB prior, the FoM drops $\sim$50--60\% for LSST and a similar trend is observed where the FoM decreases further with added intergalactic dust.
It also appears that the uncertainty added from this additional systematic is more impactful for $\Omega_\textrm{M}$ than for $w$ for LSST cosmology.
If we are able to utilize \textit{Roman}'s tighter constraint on intergalactic dust (i.e., using M10$_\textrm{med}$ rather than M10$_\textrm{max}$), LSST FoM reductions due to intergalactic dust are slightly improved, with FoM reductions at $\sim$50\% rather than $\sim$60\%.

With \textit{Roman}, the additional systematic uncertainty relative to the statistical uncertainty on individual cosmological parameters is typically comparable to those from other surveys (e.g., i.g.~dust(M10$_\textrm{max}$), SN only, 43.4\% for $w$ and 90.0\% for $\Omega_\textrm{M}$ for \textit{Roman} as compared to 32.7\% and 22.3\% for DES and 66.6\% and 106.1\% for LSST 10yr).
The reduction in FoM for \textit{Roman} with added intergalactic dust is more pronounced ($\sim$15--30\%) as compared to similar amounts of intergalactic dust with DES, but this may be expected with future surveys with tighter overall constraints on cosmology.

\subsubsection{$w_0w_a$CDM}
Finally, in $w_0w_a$CDM (Table~\ref{tab:w0wacdm}), shifts in cosmological parameters are almost all subdominant to the increase in added systematic uncertainty from each intergalactic dust variant across the three surveys except for in a few places for LSST.
Intergalactic dust results in the largest reductions in FoM for DES and LSST, with, at maximum, an approximate 25\% reduction in FoM with the nominal amount of intergalactic dust used in this work.
Although specific increases in uncertainties between cosmological parameters (i.e., $\sigma_{w_0}$, $\sigma_{w_a}$, and $\sigma_{\Omega_\textrm{M}}$) are likely degenerate, DES and LSST results demonstrate large increases for $\sigma_{\Omega_\textrm{M}}$ ($\sim$30--60\%) relative to statistical uncertainty alone.

For all surveys, FoM reductions when including intergalactic dust are less pronounced for $w_0w_a$CDM than for $w$CDM. 
For example, our results indicate more minor reductions in the FoM for LSST of $\sim$15\% with a CMB prior and 20--25\% for SNe only (for both LSST 1yr and LSST 10yr) in $w_0w_a$CDM as compared to 50--60\% reductions for all variants in $w$CDM.
This could be due to the fact that the parameter space for $w_0w_a$CDM is much less well constrained than $w$CDM, and so additional systematic uncertainties, like that of intergalactic dust, do not impact the FoM as significantly.

Finally, for \textit{Roman}, it can be observed that in $w_0w_a$CDM, cosmological parameters are more flexible in $w_0w_a$-space than for $\Omega_\textrm{M}$, resulting in little to no shifts in $\Omega_\textrm{M}$ and larger changes to the central values and uncertainties on both $w_0$ and $w_a$ with added intergalactic dust.
Specifically, with \textit{Roman}, biases and increases in uncertainty for $w_a$ are larger than those for $w_0$, but relative to the statistical uncertainty for each parameter, the increases due to intergalactic dust as a systematic are comparable to each other, approximately 20--35\% for $w_0$ and 10--20\% for $w_a$ for SNe only and somewhat larger with a CMB prior also included.

\afterpage{
\begin{rotatetable*}
\centerwidetable
\begin{deluxetable*}{l@{\extracolsep{\fill}}cccccccccccr}
\tablewidth{\textwidth}
\tablecaption{Cosmological constraints with $w_0w_a$CDM.}\label{tab:w0wacdm}
\tablehead{\colhead{Survey} & \colhead{Dust Constraint} & \colhead{Cosmo.~Const.} & \colhead{$\Delta w_0$} & \colhead{$\sigma_{\delta w_0}$} & \colhead{$\frac{\sigma_{\delta w_0}}{\sigma_{\textrm{stat},w_0}}$} & \colhead{$\Delta w_a$} & \colhead{$\sigma_{\delta w_a}$} & \colhead{$\frac{\sigma_{\delta w_a}}{\sigma_{\textrm{stat},w_a}}$} & \colhead{$\Delta \Omega_\textrm{M}$} & \colhead{$\sigma_{\delta \Omega_\textrm{M}}$} & \colhead{$\frac{\sigma_{\delta \Omega_\textrm{M}}}{\sigma_{\textrm{stat},\Omega_\textrm{M}}}$} & \colhead{$\frac{\Delta\textrm{FoM}}{\textrm{FoM}}$}\vspace{1mm}}
\startdata
\multirow{2}*{DES} & i.g.~dust(M10$_\textrm{max}$)
 & \multirow{2}*{SN only} & $-$0.0166 & 0.0403 & 34.3\% & 0.0359 & 0.2782 & 24.1\% & 0.010 & 0.020 & 30.7\% & $-$8.1\%\\
& i.g.~dust(M10$_\textrm{max}$) $\times$ 5 
 & & $-$0.0075 & 0.0473 & 40.2\% & 0.0216 & 0.2120 & 18.3\% & 0.002 & 0.033 & 51.1\% & $-$9.5\%\\
\hline
\multirow{2}*{LSST 1yr} & \multirow{2}*{i.g.~dust(M10$_\textrm{max}$)} & CMB prior & $-$0.0405 & 0.0189 & 54.1\% & 0.2808 & 0.1014 & 62.8\% & 0.010 & 0.004 & 60.1\% & $-$17.3\%\\
 & & SN only & $-$0.0419 & 0.0531 & 59.8\% & 0.0731 & 0.1689 & 50.5\% & 0.031 & 0.028 & 29.2\% & $-$26.8\%\\
\hline
\multirow{2}*{LSST 10yr} & \multirow{2}*{i.g.~dust(M10$_\textrm{max}$)} & CMB prior & $-$0.0163 & 0.0103 & 55.6\% & 0.1070 & 0.0603 & 59.1\% & 0.004 & 0.003 & 66.3\% & $-$15.5\%\\
 & & SN only & $-$0.0218 & 0.0382 & 41.3\% & $-$0.0239 & 0.2279 & 52.9\% & 0.017 & 0.032 & 32.3\% & $-$21.2\%\\
\hline
\multirow{4}*{\textit{Roman}} & \multirow{2}*{i.g.~dust(M10$_\textrm{med}$)} & CMB prior & $-$0.0022 & 0.0217 & 28.8\% & 0.0079 & 0.0797 & 32.8\% & 0.000 & 0.000 & 0.0\% & $-$5.2\%\\
& & SN only & $-$0.0001 & 0.0163 & 18.7\% & 0.0004 & 0.0297 & 9.1\% & 0.000 & 0.000 & 0.0\% & $-$2.3\%\\
& \multirow{2}*{i.g.~dust(M10$_\textrm{max}$)} & CMB prior & $-$0.0175 & 0.0331 & 43.9\% & 0.0631 & 0.1160 & 47.7\% & 0.000 & 0.000 & 0.0\% & $-$9.8\%\\
& & SN only & $-$0.0082 & 0.0300 & 34.6\% & 0.0103 & 0.0598 & 18.3\% & 0.002 & 0.000 & 0.0\% & $-$7.3\%\\
\enddata
\end{deluxetable*}

\end{rotatetable*}
}

\section{Discussion}\label{sec:discussions}

Previous SN analyses have considered intergalactic dust \citep[e.g.,][]{Corasaniti06,Goobar18,RubinUNITY,Sharma26,Dhawan26}, but none have considered using SN color evolution to constrain the signal from intergalactic dust itself.
Additionally, though \citet{RubinUNITY} propagate the impact from intergalactic dust through to cosmology, we dive into greater detail on the impact intergalactic dust has as a systematic for multiple cosmologies for upcoming data sets from both LSST and \textit{Roman}.

The DES data have at least some constraining power in outlining the true SN color distribution as a function of redshift and indicate that no more than five times the nominal amount of intergalactic dust used in this analysis (M10$_\textrm{max}$) is likely. 
Even with more constraining power and tighter constraints from LSST, we find that LSST will be unable to distinguish between no intergalactic dust and M10$_\textrm{max}$.
That being said, \textit{Roman} will be very well equipped to constrain the SN color distribution as a function of redshift as it will observe out to $z\sim3$.
LSST, therefore, will need \textit{Roman} to constrain intergalactic dust as a systematic for cosmology, and it cannot do so alone.
While it may have been expected that by using \textit{Roman}'s stricter constraint on intergalactic dust (i.e., going from M10$_\textrm{max}$ to M10$_\textrm{med}$) would have improved LSST's FoM degradation by more than just $\sim$10\%, these intergalactic dust parameterizations are not markedly different by $z=1$, and so their degradation as a systematic for LSST cosmology is similar.

Since the release of the DES5YR analysis \citep{DES24,Vincenzi24}, the calibration and analysis have been updated and improved upon in \citet{Popovic25} and \citet{Popovic26}, which we do not incorporate here given that SN color evolution is not impacted by these updates.
Similarly, the shifts we observe in cosmological parameters or increases in systematic uncertainties are insensitive to the intrinsic scatter model we choose as our baseline \citep{Popovic23} since we use the same scatter model throughout our simulations and observe the effect from adding intergalactic dust alone.
Additionally, none of the intrinsic scatter models have explicit SN color built in.

We grant that when we use SN color evolution to constrain intergalactic dust here in this work, a complicating factor may be potential color evolution intrinsic to SNe Ia themselves \citep[e.g.,][]{Foley12,Wang25}. Here, we work under the assumption that if the SN color evolution follows the effects of intergalactic dust as a function of redshift (i.e., largely volumetric), we attribute such an effect to dust. However, future analysis should continue to consider this potential degeneracy and take steps to distinguish between the two more fully, including testing for a potential $\beta$ evolution of the dust itself.

The LSST and \textit{Roman} survey strategies may be updated \citep[e.g.,][]{Rubin25}.
We expect this would result in minor changes to our analysis here in terms of constraining intergalactic dust, as the expected counts per redshift bin provided here are not expected to change by any significant factor for either survey.
As for limiting FoM reductions, LSST will have the advantage of using \textit{Roman} to constrain the impact from intergalactic dust.
And while this method of using SN color evolution to constrain intergalactic dust with \textit{Roman} may only provide a $\sim$10\% mitigation in FoM reduction for LSST, there are other avenues for constraining the distribution of intergalactic dust that can be used in combination with this method.

More work is being done to find novel methods for uncovering the distribution of intergalactic dust in our universe which could mitigate the impact intergalactic dust will have on cosmological parameter estimation. For example, \citet{McCleary26} use a maximum-likelihood estimator to measure the extinction experienced by light from distant luminous red galaxies through foreground galaxies. 
\hbox{\citet{Crenshaw26}} use data from Rubin's Data Preview 1 \citep{RubinDP1} to stack galaxy colors and construct a chromatic reddening profile in order to measure circumgalactic dust.

Future analyses could improve upon this analysis and other previous analyses by incorporating intergalactic dust not just as a flat component across the sky (as done in this paper), but by introducing a spatial component across the sky.
This can be done potentially by consulting various large-scale structure maps and estimating how many halos light from individual SNe Ia have passed through.
Since near-infrared light is less affected by dust and ultraviolet light is more sensitive to dust, impacts from intergalactic dust can be further probed by looking both to the near-infrared \citep[e.g.,][]{Peterson23,Peterson24} and to the ultraviolet. 

Another potential avenue for further constraining intergalactic dust using SNe is to do a careful analysis of how SN counts themselves drop off as a function of redshift. An almost complete understanding of survey selection effects and SN rates themselves may be necessary to carry out such an analysis, due to the competing effects they may present, but different amounts of intergalactic dust should produce a signal in the selection functions observed at the high-redshift end of any given SN survey. Such a constraint could be taken into account in conjunction with the methods used in this work.

\section{Conclusions}\label{sec:conclusions}

By observing how SN color evolves with redshift, we demonstrate, with real data, that we can constrain the distribution of intergalactic dust present in our universe.
From the DES SN data, we set a loose upper limit on the amount of intergalactic dust, finding there is at most five times the maximum curve presented in \citet{Menard10}, which can be ruled out at a tentative $\sim$2.6$\sigma$ significance.
While current data do not provide a tight constraint on intergalactic dust,
even with many more SNe observed in a simulated sample of the LSST survey, we expect that a measurement of intergalactic dust from LSST will only be marginally more constraining than DES, at $\sim$1.7$\sigma$ when comparing no intergalactic dust to the nominal amount of intergalactic dust used in this analysis, M10$_\textrm{max}$.
Only with \textit{Roman}, with its much longer redshift lever arm, do we expect to place a quality constraint on intergalactic dust, with a significance of $\sim$3.6$\sigma$ compared to the nominal amount of intergalactic dust.

We then project the impact from intergalactic dust as a systematic on cosmological parameter measurements for each survey and find that it results in up to a 0.02 mag shift in distance modulus values at $z\gtrsim1$ and significant reductions in FoM values.
While our analysis indicates that intergalactic dust does not bias cosmological parameter measurements for DES significantly, DES likely should have included intergalactic dust as a systematic, as we find it to be a large systematic with a 16.9\% and 8.1\% FoM degradation for DES in $w$CDM and $w_0w_a$CDM, respectively.

From LSST simulations (both 1yr and 10yr), we observe $\sim$60\% reductions in $w$CDM FoM values when including the nominal amount of intergalactic dust used in this analysis as a systematic.
Given that we find \textit{Roman} will be able to set a tighter constraint on the distribution of intergalactic dust than any previous survey, by using such a constraint from \textit{Roman}, we determine that LSST will observe a somewhat mitigated FoM degradation at $\sim$50\% $w$CDM FoM reductions.
We estimate intergalactic dust results in a $\sim$20\% reduction in FoM for LSST in $w_0w_a$CDM.

Lastly, with \textit{Roman}, we find that intergalactic dust will result in $\sim$20\% reductions in the FoM for $w$CDM and $\sim$5\% for $w_0w_a$CDM.
Because \textit{Roman} will observe out to $z\sim3$, it will be able to characterize intergalactic dust extremely well, and its cosmological results will not be as impacted as other surveys.
And although added intergalactic dust most often results in more negative $w_0$ values and more positive $w_a$ values (moving current values of $w_0$ and $w_a$ closer to Flat-$\Lambda$CDM), the shifts are all minor and are accompanied by comparable increases in systematic uncertainty.
Therefore, these findings do not explain away recent hints pointing towards evolving dark energy.

\begin{acknowledgements}

This paper has undergone internal review in the LSST Dark Energy Science Collaboration.
The internal reviewers were Bruno Sánchez and Lluís Galbany.
This is also a Roman Science Collaboration (RSC) publication, prepared and reviewed in accord with the RSC publication policy. This research has made use of NASA’s Astrophysics Data System.

We would like to acknowledge both Dragan Huterer and Ryan Foley for helpful discussions and suggestions for this work.
D.B.~is supported by Department of Energy grant DE-SC0016021 and the Templeton Foundation.
D.S.~is supported by Department of Energy grant {DE-SC0010007}, the David and Lucile Packard Foundation, the Templeton Foundation and Sloan Foundation. 
L.G.~acknowledges financial support from MCIN and AEI 10.13039/501100011033 under projects PID2023-151307NB-I00, CEX2020-001058-M, and by the MaX-CSIC Excellence Award MaX4-SOMMA-ICE.

Contribution Statements: E.R.P.~led the simulations, analysis, and paper writing for the project.
D.B.~advised the project throughout and assisted in paper writing.
D.S.~assisted in advising the project throughout and provided detailed comments on the paper.
B.C.~assisted in constructing simulations and managing the LSST host libraries.
R.C.~assisted in simulation construction and provided general science framework support throughout.
R.K.~assisted with \texttt{SNANA} simulations throughout.
B.P., B.M.R., and M.V.~assisted in simulation construction.

\section*{Software}
\texttt{SNANA} \citep{Kessler09,Kessler19}, {astropy} \citep{astropy:2013,astropy:2018},
{matplotlib} \citep{Hunter07},
{numpy} \citep{numpy11}, {PIPPIN} \citep{Pippin}.
\end{acknowledgements}

\newpage
\bibliographystyle{aasjournalv7.1}
\bibliography{main}{}

\appendix

\section{Coefficients for M10 Curve Reproduction}\label{appendix:M10_reproduction}

We report the coefficients we use in \texttt{SNANA} to reproduce the curves seen in M10's figure~9 for our intergalactic dust variants in Table~\ref{tab:M10_coefficients}. Polynomials are required to go through the origin.

\begin{table}[!h]
\centering
\caption{Coefficients used to reproduce M10 figure~9 following a simple polynomial with $A_{V, \textrm{i.g.dust}}(z) = C_1z^2+C_2z$.}\label{tab:M10_coefficients}
\begin{threeparttable}
\begin{tabularx}{0.6\columnwidth}{l@{\extracolsep{\fill}}ccc}
\hline \hline
Intergalactic Dust Variant & M10 Curve & $C_1$ & $C_2$ \\
\hline
M10$_\textrm{max}$ & Dotted Cyan & 0.03112 & 0.00560\\
M10$_\textrm{med}$ & Solid Cyan & 0.01228 & 0.01686\\
M10$_\textrm{min}$ & Solid Dark Blue & $-$0.00279 & 0.02216\\
\hline
\end{tabularx} 
\end{threeparttable}
\end{table}

\section{Additional SN Color Evolution Variations}\label{appendix:extra_c_vs_z}
\begin{figure}[!h]
    \centering
    \includegraphics[width=0.55\columnwidth]{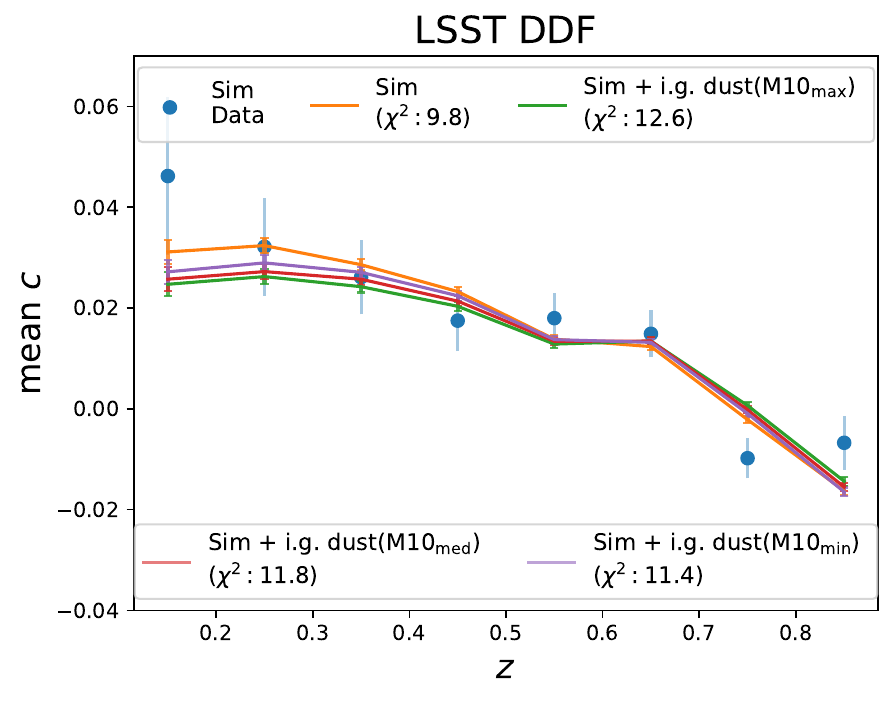}

    \caption{Similar to Fig.~\ref{fig:c_evol_LSST}, but binned mean color values as a function of redshift for LSST simulated data in the LSST deep fields with all variations of intergalactic dust curves from M10 included.}
    \label{fig:c_evol_LSST_vars}

\end{figure}

Although LSST will observe orders of magnitude more SNe Ia than DES, it will not be complete to much higher in redshift at all. To tightly constrain intergalactic dust by analyzing the evolution of SN color with redshift, a longer redshift lever arm, such as that which we expect to obtain with \textit{Roman}, is necessary. To further detail this point, we present the SN color evolution curves from all of the intergalactic dust variations provided in Fig.~\ref{fig:c_evol_ROMAN} for \textit{Roman} here for LSST as well in Fig.~\ref{fig:c_evol_LSST_vars}.
In comparing their respective $\chi^2$ values, we see that the various amounts of intergalactic dust are virtually indistinguishable from each other ($\Delta\chi^2 \lesssim 1$) and from no intergalactic dust ($\Delta\chi^2 < 3$).

\begin{figure}[!h]
    \centering
    \includegraphics[width=0.55\columnwidth]{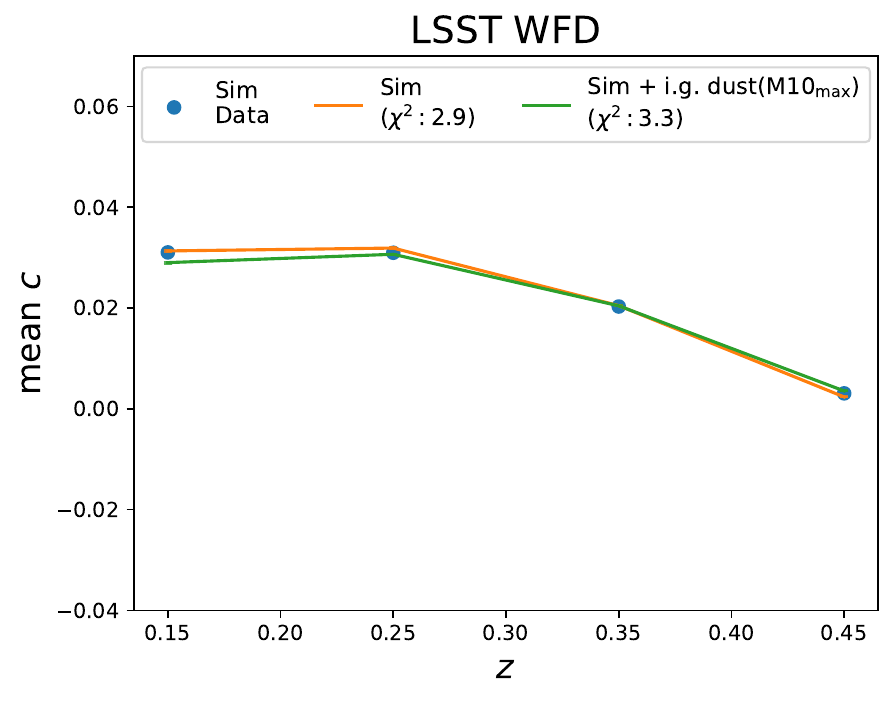}

    \caption{Similar to Figs.~\ref{fig:c_evol_LSST} and~\ref{fig:c_evol_LSST_vars}, but binned mean color values as a function of redshift for LSST simulated data in LSST's shallow, WFD fields.}
    \label{fig:c_evol_LSST_shallow}

\end{figure}

In Fig.~\ref{fig:c_evol_LSST_shallow}, even when we consider the truly unprecedented statistics of the shallow fields from LSST (which will truly have orders of magnitude more SNe Ia than DES), intergalactic dust is not expected to have enough of an impact to produce a distinguishable effect between intergalactic dust and no intergalactic dust ($\Delta\chi^2 = 0.4$).

\begin{figure}[!t]
    \centering
    \includegraphics[width=0.6\columnwidth]{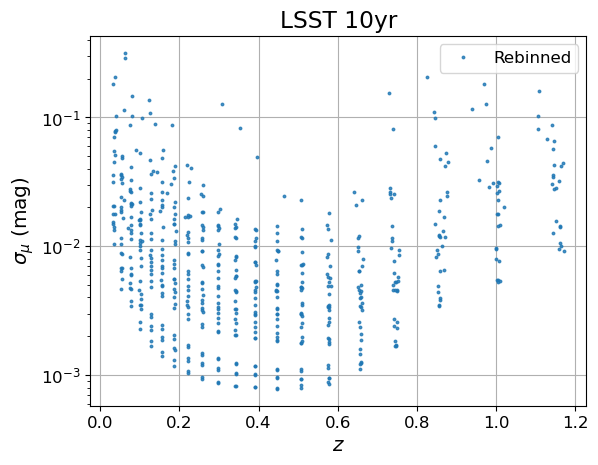}

    \caption{Distribution of rebinned distance modulus uncertainties as a function of redshift for LSST 10yr.}
    \label{fig:LSST10yr_low_muerr}

\end{figure}

\section{Uncertainty Floor Due to Extraordinarily Large Bins}\label{appendix:low_muerr}

Due to extremely high SN counts in the middle redshift bins ($0.4\lesssim z\lesssim0.6$) of our LSST simulations (both 10yr and 1yr), binned distance modulus uncertainties reached so low ($\ll$10$^{-2}$ mag, see Fig.~\ref{fig:LSST10yr_low_muerr}) that these bins dominated the cosmological fit and biased our cosmological parameter measurements.
For LSST 10yr in $w$CDM with SNe only, we obtained $w=-0.94 \pm 0.01$ ($\sim$6$\sigma$ away from the true simulated value of $w=-1)$. 
As can be seen from Fig.~\ref{fig:LSST10yr_low_muerr}, the binned distance modulus uncertainties roughly between $0.4<z<0.6$ reach an order of magnitude lower than bins on the low and high redshift ends.
These bins, in turn, dominate the cosmological fit, and although we may not expect this to result in a cosmological bias, we observe such an effect notwithstanding.
We were able to eliminate the bias observed in all of our LSST simulations by including an error floor of 10$^{-2}$ mag per bin (using the \texttt{-snrms} option in \texttt{SNANA}).
While we recognize this to be a somewhat arbitrary value, we note that with a full treatment of systematics (as will be done for a full LSST cosmology analysis), binned distance modulus uncertainties are not expected to reach below this threshold.
We deem this method to be a temporary solution, but we encourage future work that will uncover the fundamental cause of this bias and offer a rigorous solution to this problem.
Following this method we achieved unbiased results for our LSST simulations; for example, for LSST 10yr in $w$CDM with SNe only, $w=-1.00\pm0.02$.

While the statistical uncertainties observed on cosmological parameter values for both LSST 1yr and LSST 10yr appear to be similar in Tables~\ref{tab:flcdm} and~\ref{tab:wcdm}, this is due to the uncertainty floor added to both samples which dominates that effect. In this analysis, we focus on the results from LSST 1yr, while also reporting results from LSST 10yr, and leave understanding this effect as well as the true impact from intergalactic dust to the intervening time between the release of these respective data sets.

\end{document}